\documentstyle[12pt]{article}
\begin{document}
\renewcommand{\baselinestretch}{1.5}
\begin{flushright}
IP/BBSR/95-26 \\
hep-ph/9503496 \\
\end{flushright}
\begin{center}
{\bf PHASE TRANSITION IN THE EARLY UNIVERSE AND CHARGE QUANTIZATION}
\end{center}
\vspace{1in}
\begin{center}
{\bf AFSAR ABBAS} \\
{Institute of Physics} \\
{Bhubaneswar-751005, India} \\
{E-mail : afsar@iopb.ernet.in}
\end{center}
\vspace{1.5in}
\begin{center}
{\bf Abstract}
\end{center}

It is shown, for the first time , that surprisingly the
electric charge loses all physical meaning above the electro-weak phase
transition temperature. Implications of this discovery in the context
of the early
universe within the framework of various unified models
are discussed.

\vspace{1.5in}

\newpage

As the universe expands, it is predicted
that it undergoes a series of phase transitions\cite{re1,re2} during which
the appropriate symmetry breakes down in various stages until it
reaches the stage of the electro-weak (EW) symmetry.
This is given by the group structure  $SU(2)_{L}$
$\otimes$ $U(1)_{Y} $. After $ t \sim 10^{-10} $ seconds
( at  $T\sim 10^{2}$ GeV ) the electro-weak phase
transition to $ U(1){em}$
through the Higgs Mechanism takes place. It is the strucure of this phase
transition that I look into in this paper.
The conclusions drawn here are expected to have a basic and significant
impact on the whole early universe scenario including the concept of
inflation \cite{re3,re4,re5,re6} which is playing such an important role in the
present day cosmological scenarios.

The idea of the electro-weak phase transition \cite{re1} is that
above some critical temperature $T_{c}^{EW} $ the full electro-weak symmetry
$ SU(2)_{L} $$ \otimes$ $ U(1)_{Y} $ is restored. This restoration
implies that now the $ SU(2)_{L} $ gauge particles $ W^{+,-} $,
$ W^{0} $ and the $ U(1)_{Y} $ gauge particle
$ B_ {\mu} $ becomes massless. In addition
all matter particles $ e $, $\mu $, $ \tau $, u-quark etc becomes
massless too. Any model which is expected to hold at temperatures
higher than $ T_{c}^{EW} $ (e.g. GUTs, Supergravity, Superstring etc )
had better have this above mentioned property of the
EW symmetry. And indeed all
the currently available models are compatible with this. Let us ask
the question : Is there no other effect associated with the restoration
of the electroweak theory above $ T_{c}^{EW} $ ? Below I will point out that
indeed there is one.
I will discuss a new aspect arising from the restoration of
the EW symmetry at $ T_{c}^{EW} $ ( which has not been looked
into before this work). I will then discuss some implications of this
discovery.

Until 1989/1990 it was commonly believed that the electric charge was not
quantized in the Standard Model (SM) .
This is well documented in  review articles
and text books ( see for example ref. \cite{re7} ). This lack of charge
quantization was considered to be a shortcoming of the SM. It was then
found that when one goes beyond the SM, say within the framework of the
Grand Unified Theories ( GUTs) then the electric charge was automatically
quantized \cite{re7}. In fact this was thought to be the very first and
grand success of the GUTs concept.

In the cosmological context this means that at higher energies ( or
temperatures ) where the GUTs idea would be valid that the electric charge
is already quantized. In addition it was an experimental fact that at
very low energies the electric charge is also quantized. So at that
time it was felt that though the electric charge quantization was not
apparent in the SM it had to be put in by hand artificially \cite{re7}.

But recently it has been demonstrated that the above view was wrong and
that the electric charge is actually quantized in the
SM \cite{re8,re9,re10}. It has to be emphasized that the complete
structure of the SM with all it's ingredients ( like generation structure,
Higgs machanism, mass generation, anomaly cancellation etc ) is required
to give a rigorous and complete demonstration of the electric charge
quantisation in the SM \cite{re9,re10}.  As we shall require some of
the ideas later on in this paper let me summarize the arguments
demonstrating the existence of charge quantization in the SM.

Let us start by looking at the first generation of quarks and leptons
(u, d, e,$\nu$ )  and assign them to
$SU(3)_{c} \otimes SU(2)_L \otimes U(1)_Y$ representation as follows
\cite{re9,re10}.

\begin{displaymath}
q_L = \pmatrix{u \cr d}_L, (3,2,Y_q)
 \end{displaymath}
\begin{displaymath} u_R; (3,1,Y_u) \end{displaymath}
\begin{displaymath} d_R; (3,1,Y_d) \end{displaymath}
\begin{displaymath} l_L =\pmatrix{\nu \cr e}; (1,2,Y_l)
\end{displaymath}
\begin{equation}
 e_R; (1,1,Y_e)
\end{equation}

To keep things as general as possible this brings in five unknown
hypercharges.

Let us now define the electric charge in the most general way in
terms of the diagonal generators of $SU(2)_L \otimes U(1)_Y$ as
\begin{equation} Q'= a'I_3 + b'Y \end{equation}
\newline We can always scale the electric charge once as $Q={Q'\over
a'}$ and hence ($b={b'\over a'}$)
\begin{equation} Q = I_3 + bY \end{equation}

 In the SM $ SU(3)_{c} $ $\otimes$ $ SU(2)_{L}$ $\otimes$
$U(1)_{Y}$ is spontaniously broken through the Higgs mechanism to the
group $ SU(3)_{c} $ $\otimes$ $U(1)_{em}$ . In this model the Higgs is
assumed to be doublet $ \phi $ with arbitrary hypercharge $ Y_{\phi}$.
 The isospin $I_3 =- {1\over2}$ component of the
Higgs develops a nonzero vacuum expectation value $<\phi>_o$. Since we want
the $U(1)_{em}$ generator Q to be unbroken we require $Q<\phi>_o=0$. This
right away fixes b in (3) and we get
\begin{equation} Q = I_3 + ({1 \over 2Y_\phi})Y \end{equation}

Next one requires that the fermion masses arise through Yukawa coupling
and also by demanding that the triangular anomaly cancels (to ensure
renormaligability of the theory) ( see \cite{re9,re10} for details);
one obtaines all the unknown hypercharge in terms of the unknown Higgs
hypercharge $Y_{\phi}$. Ultimately $ Y_{\phi} $ is cancelled out
and one obtain the correct charge quantization as follows.

\begin{displaymath}
 q_L = \pmatrix{u \cr d}_L , Y_q = {{Y_\phi} \over{3}},  \end{displaymath}
\begin{displaymath} Q(u) = {2\over 3}, Q(d) = {-1\over 3}
 \end{displaymath}
\begin{displaymath} u_R, Y_u = {3\over{4}} {Y_\phi}, Q(u_R) ={2\over{3}}
 \end{displaymath}
\begin{displaymath} d_R, Y_d = {-2\over{3}} {Y_\phi}, Q(d_R) ={-1\over3}
 \end{displaymath}
\begin{displaymath} l_L = \pmatrix{\nu \cr e}, Y_l = -Y_\phi,
Q(\nu) = 0, Q(e) = -1
  \end{displaymath}
\begin{equation}
 e_R, Y_e = -2Y_\phi, Q(e_R) = -1
\end{equation}

It has been shown \cite{re9} that  for arbitrary $ N_{c} $ the colour
dependence of the electric charge as demanded by the SM is

\begin{displaymath}
\newline Q(u) = {1\over 2}(1+{1\over N_c})
\end{displaymath}
\begin{equation}
\newline Q(d) = {1\over 2}(-1+{1\over N_c})
\end{equation}

The implication of this for the Grand Unified Theories has been discussed
in ref \cite{re11}.

 One should note that  equations  (5) and (6)
show that contrary to all earlier expectations, the electric charge is
quantized in SM. The complete structure of the SM as is, is required
to obtain this result on very general grounds. The SM is the best
tested model of particle physics.
As long as these assumptions are valid as one goes beyond it one
should maintain all the intrinsic properties of it. What has this to
say about the early universe scenarios available today ?

Clearly the $ U(1)_{em}$ symmetry which
arose due to spontaneous symmetry breaking due to a Higgs doublet
in the EW symmetry will be lost above $ T_{c}^{EW} $ whence
$ SU(2)_{L}$ $\otimes$ $ U(1)_{em}$ symmetry would be
restored. As is obvious, above $ T_{c}^{EW}$ all the fermions and gauge bosons
becomes massless \cite{re1,re2}. This properly is well-known and
has been incorporated in cosmological models. Here I point out a new
phenomenon arising from the restoring of the full EW symmetry .

Note to start with the parameter b and Y in equation (3) in the definition
of electric charge were completely unknown. We could lay a handle on 'b'
entirely on the basis of the presence of spontaneous symmetry breaking
and on ensuring that photon was massless $ b = \frac{1}{2 Y_{\phi}} $.
Above $ T_{c}^{EW} $ where the EW symmetry is restored there is
no spontaneous symmetry breaking and hence the parameter b is
completely undetermined. Together 'bY' could be any arbitrary number
whatsoever even an irrational number. Within
the framework of this model above $ T_{c}^{EW} $ we just cannot define electric
charges for a fermion at all. It may be a number which is zero or infinite
or an irrational etc. Hence the electric charge given by equation (3)
loses any physical meaning all together.This is the new amazing result.

So above $ T_{c}^{EW} $ all the particles have not only become massless,
they have forgotten their charges also. It just does not make sense to talk
of their charges.
The concept of electric charge has been lost.
There is no such thing as charge anymore.
The photon (which was a linear combination of $W^{0}$
and $ B_{\mu} $ after spontaneous symmetry breaking ) with it's
defining  vector
characterteristic  does not exist either. So the conclusion is that there is
no elctrodynamics above $ T_{c}^{EW} $.

Note that the electric charge in the SM was not an elementary or fundamental
object at all. In fact it was
a secondary quantity defined in terms of the elementary objects
$ I_{3}$ and Y ( see equation(3)). So it should not be really surprising to
see it loose it's meaning under special circumstances. The same is true
of the photon of $ U(1)_{em} $ . In short $ U(1)_{em} $ owes its
existence to SSB and looses it's meaning when
the full EW symmetry is restored. Interestingly
the fact that people have found \cite{re12} QED to be inconsistent for
massless fermions is a puzzle no more.

As noted above earlier (prior to 1989/1990) electric charge was quantized
in GUTs and this was artificially imposed on SM.
Now we have demonstrated that actually electric charge is quantised in
SM and when at high temperature the EW symmetry is
unbroken, the concept of electric charge does not arise. At those and
still higher temperatures \cite{re1,re2,re3,re4,re5,re6} extensions
like GUTs, Supergravity,
Superstrings are believed to be relevant. Earlier electric charge quantisation
was thought to be a success of the GUTs idea but now in the light of
the new development discussed here, this becomes it's
major weakness ( note that
the GUTs idea is supposed to hold above $ T_{c}^{EW} $ ). Similarly
for the accepted extensions which have become standard in the current
cosmological scenarios this becomes a basic problem too.

In short a particular model which pertains to be valid at
temperatures higher than $ T_{c} ^{EW}$ is in trouble if it has charge
quantization built into it. It is demanding something which the SM
does not require. Clearly we have to be extremely wary when we are
trying to extend models beyond the SM
right to $ t \sim 10^{-44} $ seconds.

Clearly the fact that the electric charge loses any
physical meaning and it's very existence above
$ T_{c}^{EW} $  will have major impact on our models of the early universe.
How the presently accepted
models and scenarios will change in the light of the
new information given here has to be studied carefully.


\end{document}